\documentclass[conference]{IEEEtran}
\IEEEoverridecommandlockouts
\usepackage{cite}
\usepackage{amsmath,amssymb,amsfonts}
\usepackage{algorithmic}
\usepackage{subcaption}
\usepackage{graphicx}
\usepackage{textcomp}
\usepackage{xcolor}
\def\BibTeX{{\rm B\kern-.05em{\sc i\kern-.025em b}\kern-.08em
    T\kern-.1667em\lower.7ex\hbox{E}\kern-.125emX}}

\usepackage{enumitem}
\begin{document}

%#############################################

\title{CoRPA: Adversarial Image Generation for Chest X-rays Using Concept Vector Perturbations and Generative Models}

%\author{Anonymous Authors
%}

\author{\IEEEauthorblockN{Amy Rafferty}
\IEEEauthorblockA{\textit{School of Informatics} \\
\textit{Universiy of Edinburgh}\\
Edinburgh, UK \\
0000-0002-5120-966X}
\and
\IEEEauthorblockN{Rishi Ramaesh}
\IEEEauthorblockA{\textit{NHS Lothian} \\
Edinburgh, UK \\
}
\and
\IEEEauthorblockN{Ajitha Rajan}
\IEEEauthorblockA{\textit{School of Informatics} \\
\textit{University of Edinburgh}\\
Edinburgh, UK \\
0000-0003-3765-3075}
}

\maketitle

\begin{abstract}

Deep learning models for medical image classification tasks are becoming widely implemented in AI-assisted diagnostic tools, aiming to enhance diagnostic accuracy, reduce clinician workloads, and improve patient outcomes. However, their vulnerability to adversarial attacks poses significant risks to patient safety. Current attack methodologies use general techniques such as model querying or pixel value perturbations to generate adversarial examples designed to fool a model. These approaches may not adequately address the unique characteristics of clinical errors stemming from missed or incorrectly identified clinical features.
We propose the Concept-based Report Perturbation Attack (CoRPA), a clinically-focused black-box adversarial attack framework tailored to the medical imaging domain. CoRPA leverages clinical concepts to generate adversarial radiological reports and images that closely mirror realistic clinical misdiagnosis scenarios. We demonstrate the utility of CoRPA using the MIMIC-CXR-JPG dataset of chest X-rays and radiological reports.
Our evaluation reveals that deep learning models exhibiting strong resilience to conventional adversarial attacks are significantly less robust when subjected to CoRPA's clinically-focused perturbations. This underscores the importance of addressing domain-specific vulnerabilities in medical AI systems. By introducing a specialized adversarial attack framework, this study provides a foundation for developing robust, real-world-ready AI models in healthcare, ensuring their safe and reliable deployment in high-stakes clinical environments.

\end{abstract}


\section{Introduction}

AI-assisted diagnostic systems have emerged as a transformative technology in medical imaging, offering significant potential to enhance diagnostic accuracy, reduce workload, and improve patient outcomes. These systems leverage deep learning algorithms to analyze medical images, such as chest X-rays, CT scans, and MRIs, for the detection and classification of various pathologies with performance often comparable to or exceeding that of human experts \cite{b1, b2, b3}. The integration of AI into diagnostic workflows holds the promise of faster, more consistent interpretations, assisting radiologists in identifying early signs of disease, optimizing treatment plans, and reducing diagnostic errors \cite{b4, b5}. However, concerns regarding their robustness and interpretability highlight the need for continued research to ensure their safe deployment in high-stakes clinical settings.

The concept of adversarial attacks was first introduced in 2014 by a ground-breaking study \cite{b6} which demonstrated that small, imperceptible perturbations to input data could mislead deep learning models, causing significant misclassification rates even in cases where these perturbations were imperceptible to the human eye. This work revealed concerning vulnerabilities in neural networks and sparked significant research into adversarial machine learning and the security of AI systems. Adversarial attacks can take various forms, including white-box attacks, where the attacker has full access to the model, and black-box attacks, where the attacker only has access to the model's inputs and outputs \cite{b7}.

Numerous studies have shown that adversarial examples can effectively manipulate deep learning systems across various clinical domains \cite{b8, b9}. These attacks typically follow general-purpose approaches, such as generating small data perturbations, submitting numerous queries to estimate model parameters, or utilizing substitute models with transferability~\cite{b10}. 
Although such methods are commonly used in computer vision, they may not adequately address the unique characteristics of clinical data, and the semantic meanings of complex clinical features, leading to the generation of adversarial examples that do not accurately represent real-world scenarios such as diagnostic errors. To address this gap, we propose the development of a specialized adversarial attack framework tailored to the medical domain, that facilitates model evaluation with respect to misinterpreted or missed clinical features in images. %which would facilitate a more realistic evaluation of model vulnerabilities.

In this work, we introduce the Concept-based Report Perturbation Attack (CoRPA), a novel, clinically-focused black-box adversarial attack specifically designed for the medical imaging domain. CoRPA leverages clinical features, or concepts, associated with specific pathologies in the dataset to generate concept vectors for each image-report pair. These concept vectors are deliberately perturbed to simulate noisy, incorrectly identified, or missing clinical features in a radiograph, replicating real-world scenarios that could lead to misdiagnosis. The perturbed concept vectors are used to generate adversarial radiological reports, which are subsequently input into a text-to-image generative model to create adversarial medical images. We illustrate the CoRPA technique using MIMIC-CXR-JPG~\cite{b11, b12, b13}, a large public dataset of chest X-rays and their associated free-text radiological reports. %~\cite{b12}~\cite{b13}. 

Published statistics on the frequency of missed clinical findings in chest X-rays~\cite{b66} reveal that 43\% of malpractice claims come from common diagnostic errors such as missed lesions, opacities (eg. pneumonia), pneumothorax, and mediastinal and hilar abnormalities. These critical clinical features are fully integrated into CoRPA’s concept vectors for the MIMIC-CXR-JPG dataset. The prevalence of these errors underscores the necessity for a clinically-grounded adversarial testing framework such as CoRPA, to evaluate and enhance the robustness of AI systems against real-world challenges in medical diagnostics.

While we showcase CoRPA's application using the MIMIC-CXR-JPG dataset, its design allows seamless extension to other medical datasets containing both image and report data. CoRPA effectively bridges the gap between generic adversarial attack methods and the specific demands of medical image diagnosis.

%Although we demonstrate CoRPA's application on MIMIC-CXR-JPG , CoRPA is easily extendable to other medical datasets that contain both image and report data. CoRPA bridges the gap between general-purpose adversarial attacks and the unique requirements of healthcare applications.

We evaluate the robustness of several deep learning architectures to both clinically-focused adversarial images generated by CoRPA and adversarial images produced by general-purpose attacks. Our findings reveal that models which exhibit strong resilience to standard white-box and black-box attacks tend to demonstrate significantly lower robustness when subjected to CoRPA attacks. This highlights the importance of addressing domain-specific vulnerabilities in medical AI systems, as conventional adversarial attacks may fail to expose models to more realistic, clinically-relevant errors.

\noindent In summary, the key contributions in this paper are:
\begin{enumerate}
\item A novel adversarial testing framework, CoRPA, which generates adversarial examples that replicate incorrectly identified or overlooked clinical findings by perturbing clinical concepts in radiology reports. 
\item CoRPA introduces an innovative application of text-to-image generative models to generate adversarial radiographs derived from modified radiology reports.
\item Extensive evaluation of the robustness of seven state-of-the-art (SOTA) top-performing MICCAI challenge models for chest X-ray diagnosis, using the CoRPA adversarial examples.
\item Comparison of CoRPA attack effectiveness against three other SOTA adversarial testing techniques. 
\end{enumerate}

%By introducing a clinically-focused adversarial attack methodology, CoRPA aims to advance the development of robust ML models that can withstand real-world challenges in medical imaging. This study provides a framework for future research in adversarial robustness within the healthcare domain.

\section{Background}

Adversarial attacks exploit vulnerabilities in machine learning models by introducing subtle perturbations to input data, which are often imperceptible to humans but can lead to significant model misclassification rates \cite{b6} \cite{b14}. These attacks pose critical challenges in high-stakes applications such as medical diagnostics, where errors in AI-assisted diagnostic systems may result in substantial risks to patient safety. Such misclassifications undermine trust in the reliability and utility of automated diagnostic systems \cite{b15}.

Adversarial attacks are typically categorized based on the attacker’s knowledge of the target model and the attack's overall objective \cite{b16}. Depending on the goal, attacks can be classified as targeted, where the aim is to produce specific misclassifications, or untargeted, which aim to broadly disrupt model accuracy.

White-box attacks assume complete access to the model’s architecture, parameters, and training data. These attacks are highly effective~\cite{b17} and include several widely studied techniques in the literature, such as Basic Iterative Method (BIM)~\cite{b18}, DeepFool~\cite{b19}, and the Carlini \& Wagner (C\&W) attack~\cite{b17}. Among the most prevalent methods are the Fast Gradient Sign Method (FGSM)~\cite{b14} and Projected Gradient Descent (PGD)~\cite{b20}. FGSM calculates the gradient of the loss with respect to the input image, and generates an adversarial image by perturbing the input in the direction of the gradient sign. PGD, considered one of the strongest first-order attacks, iteratively seeks perturbations that maximize the model's loss while constraining their magnitude within a predefined limit.

In contrast, black-box attacks assume that the attacker has no access to the target model's internal details and largely rely on external methods and approximations to craft adversarial examples. These attacks are much more realistic - in a real-world environment, especially for a medical diagnostic model, it is very unlikely that an attacker will have access to the model's internal information~\cite{b8}. These include transfer-based techniques such as Iterative Fast Gradient Sign Method (I-FGSM)~\cite{b18} and Skip Gradient Method (SGM)~\cite{b21}, which leverage the transferability property by using a surrogate model to generate adversarial examples, and score-based techniques such as Simple Black-Box Attack (SimBA)~\cite{b22} and Natural Evolution Strategy (NES)~\cite{b23}, which estimate gradients through repeated model queries.

Given the significant threat posed by adversarial examples, extensive research has been devoted to developing defense mechanisms to mitigate their impact. Adversarial training~\cite{b20} involves fine-tuning the model on adversarial examples to enhance its robustness. Input transformations~\cite{b24} aim to reduce the noise introduced by adversarial perturbations before the data is processed by the model. Randomization strategies~\cite{b25} add stochasticity to the model's inference process, making it more resilient to adversarial attacks. Model ensembles~\cite{b26} leverage multiple models with diverse architectures to improve overall robustness by reducing the likelihood of all models being simultaneously compromised.

In medical diagnostics, adversarial attack and defense strategies are extensively applied to medical image classification models to evaluate and enhance their robustness~\cite{b10}. However, these methods are predominantly developed within a general computer vision framework, and have no clinical context for what they are perturbing. Therefore, the generated adversarial examples may fail to accurately represent realistic threats that a model could encounter in a clinical setting.

This study introduces CoRPA, a clinically-focused novel black-box untargeted attack that leverages a text-to-image generative model and concept vector perturbations to generate adversarial images for a dataset of chest X-rays and linked radiological reports. These images are synthesized based on adversarial reports, which are constructed through perturbations of concept vectors, as described in Section~\ref{corpa}.

\subsection{Applications of Clinical Concept Vectors}

Concept vectors associated with medical images allow us to represent specific semantic or clinical features within a multidimensional space. For example, they can encode the presence of clinically significant attributes such as \texttt{mass} or \texttt{enlarged heart} in chest X-rays. The direct relationship between these vectors and their corresponding medical images facilitates the interpretation and manipulation of data in a way that is interpretable to humans as well as deep learning models.

In recent years, concept vectors have been extensively applied in the literature across several key domains related to medical imaging, including the interpretability and robustness of SOTA classification models \cite{b27}.

In the field of Explainable Artificial Intelligence (XAI), which seeks to overcome the trustworthiness and transparency challenges associated with the `black-box' nature of many SOTA deep learning models~\cite{b28, b29}, concept vectors have gained prominence in explanation techniques due to their human-readable properties. Explanation methods such as Concept Bottleneck Models (CBMs)~\cite{b30} leverage clinical concepts to introduce a fully interpretable intermediate step to the traditional classification pipeline, offering users accessible interpretations of model decisions. Extensive research has focused on the automatic generation of concept vectors for images~\cite{b31,b32,b33}, addressing the challenge posed by the lack of publicly available concept-annotated datasets. This has led to the development of several high-performing concept-based explanation techniques for medical image classification models~\cite{b34, b35, b36}, such as Cross-Modal Conceptualization in Bottleneck Models (XCBs)~\cite{b37}, and Automated Concept-based Explanation (ACE)~\cite{b38}.

Concept vectors have also been used in recent studies to enhance the robustness of SOTA classification models. By linking input images to meaningful concept vectors representing image features, models can learn to detect and mitigate adversarial manipulations. In a recent study~\cite{b39}, a method for detecting adversarial examples using high-level Concept Activation Vectors (CAVs) is introduced. CAVs capture human-interpretable concepts in neural network activations and help identify deviations in model behaviour, making adversarial examples detectable.

To the best of our knowledge, no studies have yet introduced methods of using clinical concept vectors to attack a model—such methods have primarily been explored for adversarial defence and detection  purposes. In contrast, our approach (CoRPA) leverages these concept vectors, along with their perturbations, to generate adversarial radiological reports. These reports are subsequently input into a text-to-image generative model to produce realistic adversarial medical images.

\subsection{Chest X-ray Generation Using Generative Models}

Text-to-image generative models, such as Generative Adversarial Networks (GANs) \cite{b40} and Diffusion models~\cite{b41}, enable the synthesis of realistic images from unstructured input text. These models have been applied in recent research to support medical data augmentation and generation~\cite{b42}, addressing the challenge of the limited availability of diverse, high-quality public medical datasets. A recent study by Stanford University~\cite{b43} demonstrated the implementation of a Stable Diffusion model~\cite{b44} to automatically generate high-quality chest X-rays with specified abnormalities. Numerous studies have reported notable success in generating chest X-rays using fine-tuned Diffusion models~\cite{b45}, Stable Diffusion models~\cite{b46}, and GANs~\cite{b47}.

For generating adversarial chest X-rays within our proposed attack framework CoRPA, we employ a Stable Diffusion model owing to its substantially lower computational requirements compared to other generative models. Stable Diffusion operates within a reduced-dimensional latent space rather than the high-resolution pixel space of the input image.

\section{Materials}

This section introduces the dataset utilized in this study, including the pre-processing and label annotation methodologies used. It also provides an overview of the classification model architectures whose robustness we examine, evaluated through both our proposed untargeted black-box adversarial technique (CoRPA) and other commonly-used attack methods.

\subsection{Dataset}

We use the public anonymized MIMIC-CXR-JPG dataset from PhysioNet~\cite{b11,b12,b13}, consisting of chest X-rays and corresponding free-text radiological reports. The dataset is pre-processed to include only chest X-rays with associated reports. To minimize confounding variables~\cite{b48}, we focus exclusively on images acquired from the standard Posteroanterior (PA) viewpoint, excluding alternative perspectives such as Anteroposterior (AP) and lateral views. Following this filtering process, the dataset consists of 85,872 unique image-report pairs. The images are resized to 512 pixels to optimize computational efficiency and address storage constraints.

Although MIMIC-CXR-JPG provides pathology label annotations, generated automatically through both CheXpert \cite{b49} and NegBio \cite{b50}, these NLP-based labels have been found to be unreliable \cite{b51} \cite{b52}. We instead label the dataset ourselves using clinical concept vectors; the approach for this is detailed in Section~\ref{labelling}. Based on these pathology annotations, we further filter our dataset to contain only image-report pairs belonging to the following labels, derived from the original MIMIC-CXR-JPG label set: Healthy (No Finding), Cancer (Lung Lesion), Cardiomegaly, Pleural Effusion, Pneumonia and Pneumothorax. These six pathology labels were selected from the original set of fourteen in the MIMIC-CXR-JPG dataset under radiologist guidance. The selection criteria included their clinical significance in real-world diagnostic practice and their representation of general pathological conditions rather than specific symptoms (e.g., Consolidation, Lung Opacity).

The resulting dataset has a significant class imbalance, with the majority of cases labelled as Healthy. To reduce the impact of this imbalance on the performance of classification models, we apply the One-Sided Selection \cite{b53} undersampling technique. The dataset is then divided into training, validation, and testing sets using an 80/10/10 split ratio. Class frequencies within each subset are presented in Table~\ref{freqs}.

\begin{table}[b]
\caption{Pathology label Frequencies in our dataset}
\begin{center}
\begin{tabular}{|c|c|c|c|}
\hline
\textbf{Label}&\textbf{Training}&\textbf{Validation}&\textbf{Testing} \\
\hline
Healthy & 12963 & 1620 & 1620 \\
Cancer & 1143 & 143 & 143 \\
Cardiomegaly & 4590 & 573 & 574 \\
Pleural Effusion & 5942 & 743 & 743 \\
Pneumonia & 2582 & 323 & 323 \\
Pneumothorax & 1494 & 186 & 187 \\
\hline
Total & 28714 & 3588 & 3590 \\
\hline
\end{tabular}
\label{freqs}
\end{center}
\end{table}

\subsection{Dataset Labelling}\label{labelling}

\begin{table*}[t]
\caption{Clinical Concept Selection. Original report phrases are clustered to create clinical concepts for each pathology label.}
\begin{center}
\begin{tabular}{|c|c|c|}
\hline
\textbf{Label}&\textbf{Clinical Concepts}&\textbf{Original Phrases} \\
\hline
\textbf{Healthy} & Unremarkable & Normal; Unremarkable; Lungs clear$^{\mathrm{b}}$; No evidence; No interval change$^{\mathrm{b}}$; \\
& & No acute cardiopulmonary abnormality$^{\mathrm{b}}$; Normal hilar contours$^{\mathrm{b}}$; No acute process$^{\mathrm{b}}$ \\
\hline
\textbf{Cancer} & Mass & Mass; Cavitary lesion$^{\mathrm{b}}$; Carcinoma; Neoplasm; Tumor/Tumour; Rounded opacity$^{\mathrm{b}}$; Lung cancer; \\
& & Apical opacity; Lump; Triangular opacity; Malignant; Malignancy \\
\cline{2-3}
& Nodule & Nodular densities/density; Nodular opacities/opacity; Nodular opacification; Nodule \\
\cline{2-3}
& Irregular Hilum & Hilar mass; Hilar opacity; Hilus enlarged$^{\mathrm{b}}$, Hilus fullness$^{\mathrm{b}}$, Hilus bulbous$^{\mathrm{b}}$ \\
\cline{2-3}
& Adenopathy & Mediastinal lymphadenopathy; Mediastinal adenopathy; Hilar lymphadenopathy; Hilar adenopathy \\
\cline{2-3}
& Irregular Parenchyma & Pulmonary metastasis; Carcinomatosis; Metastatic disease \\
\hline
\textbf{Pneumonia} & Pneumonitis & Pneumonia; Pneumonitis; Bronchopneumonia; Airspace disease; Air bronchograms; Cavitation \\
\cline{2-3}
& Consolidation & Consolidation \\
\cline{2-3}
& Infection & Infection; Infectious process; Infectious \\
\cline{2-3}
& Opacities$^{\mathrm{a}}$ & Airspace opacities/opacity; Homogeneous opacities/opacity; Patchy opacities/opacity; \\
& & Ground-glass opacities/opacity; Alveolar opacities/opacity; Ill-defined opacities/opacity; \\
& & Reticulonodular pattern \\
\hline
\textbf{Pleural Effusion} & Effusion & Effusion; Effusions; Pleural effusion \\
\cline{2-3}
& Fluid & Pleural fluid; Fluid collection; Layering fluid$^{\mathrm{b}}$ \\
\cline{2-3}
& Meniscus Sign & Meniscus, Meniscus sign \\
\cline{2-3}
& Costophrenic Angle & Costophrenic angle blunting$^{\mathrm{b}}$ \\
\hline 
\textbf{Cardiomegaly} & Enlarged Heart & Cardiomegaly; Borderline cardiac silhouette/heart$^{\mathrm{b}}$; Prominent cardiac silhouette$^{\mathrm{b}}$; \\
& & Heart enlarged$^{\mathrm{b}}$; Top-normal heart$^{\mathrm{b}}$ \\
\hline
\textbf{Pneumothorax} & Absent Lung Markings & Absent lung markings$^{\mathrm{b}}$; Apical pneumothorax; Basilar pneumothorax; \\
& & Hydro pneumothorax/Hydropneumothorax; Lateral pneumothorax; Pneumothorax; Pneumothoraces \\
\cline{2-3}
& Irregular Diaphragm & Flattening of ipsilateral diaphragm$^{\mathrm{b}}$; Inversion of ipsilateral diaphragm$^{\mathrm{b}}$ \\
\hline
\multicolumn{3}{l}{$^{\mathrm{a}}$The concept `Opacities' can belong to Pneumonia or Pleural Effusion, depending on the presence of other concepts. This is explained in Section~\ref{labelling}.}\\
\multicolumn{3}{l}{$^{\mathrm{b}}$Type B: Word order may vary.}\\
\end{tabular}
\label{tab:concepts}
\end{center}
\end{table*}

\begin{figure}[bt]
\centerline{\includegraphics[width=0.9
\linewidth]{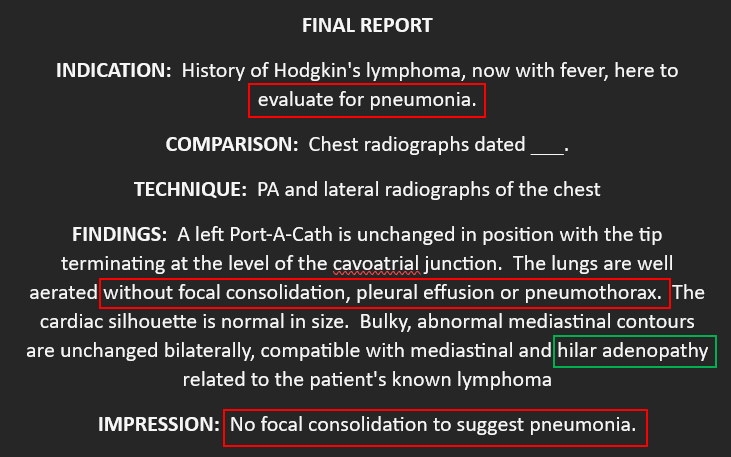}}
\caption{Example of our labelling approach outperforming CheXpert. The phrases in red were used by CheXpert, which incorrectly labelled this report as \textbf{Pneumonia}. Our approach accounts for the negative mentions and context, and based on the \textbf{Hilar adenopathy} (green), labels the report as \textbf{Cancer}.}
\label{fig:labelling}
\end{figure}

MIMIC-CXR-JPG provides NLP-generated labels from radiology reports using CheXpert \cite{b49} and NegBio \cite{b50}. However, these labels often exhibit high false-positive rates due to misinterpretation of context and negations \cite{b52}. Instead of using these, we annotate the data using clinical concept vectors. An example of our approach outperforming CheXpert is shown in Figure~\ref{fig:labelling}.

A consultant radiologist with over 15 years of experience analyzed a subset of radiological reports to identify words and phrases indicative of the six pathology labels (Healthy, Cancer, Cardiomegaly, Pleural Effusion, Pneumonia, and Pneumothorax). This resulted in a list of diagnostic phrases commonly used by experts in chest X-ray reports for the targeted pathologies. To reduce sparsity and redundancy (e.g., \texttt{nodular opacity} vs. \texttt{nodular opacities}), these phrases were clustered under the radiologist's guidance into consolidated clinical concepts for each label. The clusters and original phrases are shown in Table~\ref{tab:concepts}.

To convert a free-text radiological report into a concept vector corresponding to the presence of the 17 resulting clinical concepts within the text, we first clean the report, and then determine which of the original phrases associated with each concept are present within the formatted text.

Given the variability in report structures, and following radiologist guidance, we restrict our analysis to the FINDINGS and IMPRESSION report sections \cite{b11}, discarding sections such as HISTORY and COMPARISON, which may introduce noise. Pre-processing involves splitting the report into sentences, removing punctuation and formatting characters (e.g., newline, tab), and converting text to lowercase. From the resulting sentences, we remove the following:

\begin{itemize}
    \item Sentences with less than 2 words
    \item Sentences starting with negating phrases (“no ", “there is no ", “no evidence of”)
    \item Sentences containing false positive inducing terms (e.g. “nipple shadow” in the sentence “nodular opacity is likely a nipple shadow”, or “evaluate” in the sentence “evaluate for pneumonia”)
    \item Parts of sentences following a negating statement (e.g. “clear of”, “without”, “should not be mistaken for”)
\end{itemize}

This process produces a refined set of formatted sentences, free from negations and misleading terms, suitable for concept detection. The concept vector for each report is a binary representation, indicating the presence of each clinical concept within these formatted sentences.

As shown in Table~\ref{tab:concepts}, clinically relevant phrases for each concept can be categorized as type A, which are fully encapsulated phrases (e.g., \texttt{mass}, \texttt{pleural fluid}), or type B, which are flexible word collections that can appear in different orders (e.g., \texttt{hilus enlarged} as ``the hilus appears enlarged" or ``enlarged hilus"). For Type A phrases, the concept's presence in the report's concept vector is set to 1 if the exact phrase appears in at least one formatted sentence. For Type B phrases, each formatted sentence is checked to determine if all words in the phrase, or their synonyms (e.g., ``hilum" = ``hilus" = ``hilar" or ``heart size" = ``cardiac silhouette" = ``cardiac contour"), are present. If every word or its synonym is found in a sentence, the concept's presence is marked as 1 in the concept vector.

After this process, each report—and by extension, each chest X-ray—is assigned a concept vector indicating the presence of 17 clinical concepts. We note that the \texttt{Opacities} concept may apply to either Pneumonia or Pleural Effusion. Under radiologist guidance, we assume the presence of \texttt{Opacities} in a concept vector corresponds to Pleural Effusion, unless other concepts associated with Pneumonia (e.g., \texttt{Infection}) are present, in which case the Pneumonia label is assigned. As a final pre-processing step, the \texttt{Unremarkable} concept, representing the Healthy class, is set to 0 if any other concept is present. This prevents mislabeling cases where a report mentions a pathology in one lung but describes the other as clear, which could erroneously invoke the Healthy class. These concept vectors, directly corresponding to pathologies, are used to label the dataset. Chest X-rays associated with multiple labels are duplicated and assigned each label accordingly.

\subsection{Evaluation Models}

To implement CoRPA, our concept-based adversarial attack, we first train multiclass classification models on our dataset and evaluate their performance on both the original test set and the adversarial images generated by the attack.

We assess the robustness of the backbone model architectures used by the highest performing published submissions in the recent MICCAI challenge, CXR-LT \cite{b54} \cite{b13}, which focused on developing multiclass classification models that address class imbalances in the MIMIC-CXR-JPG dataset. The leading models from this challenge were primarily ensemble models, combining these backbone architectures. Therefore, we consider it essential to directly assess the robustness of these foundational models.

The models selected for evaluation are ResNet50 \cite{b55}, ResNet101 \cite{b55}, ResNeXt101 \cite{b56}, DenseNet161 \cite{b57}, ConvNeXt-S \cite{b58}, ConvNeXt-B \cite{b58}, and EfficientNetV2-S \cite{b59}. Each model is trained using PyTorch on an NVIDIA GTX 1060 6GB GPU, with CrossEntropyLoss and an SGD optimizer, a learning rate of 0.001, and momentum of 0.9. Training is conducted for a maximum of 20 epochs with early stopping enabled.

We evaluate model performance, as done in CXR-LT \cite{b60}, using mean Average Precision (mAP) and mean AUROC across the six classes. mAP is considered the primary metric as it is not adversely affected by class imbalance, while AUROC has been shown to be inflated in imbalanced datasets \cite{b61}. We present both metrics for thoroughness. We further assess the robustness of these models using the Attack Success Rate (ASR) of CoRPA, as well as three other widely implemented adversarial attacks, FGSM \cite{b14}, PGD \cite{b20} and SimBA \cite{b22}.

\section{CoRPA: Concept-based Report Perturbation Attack}\label{corpa}

\begin{figure*}[t]
\centerline{\includegraphics[width=
\linewidth]{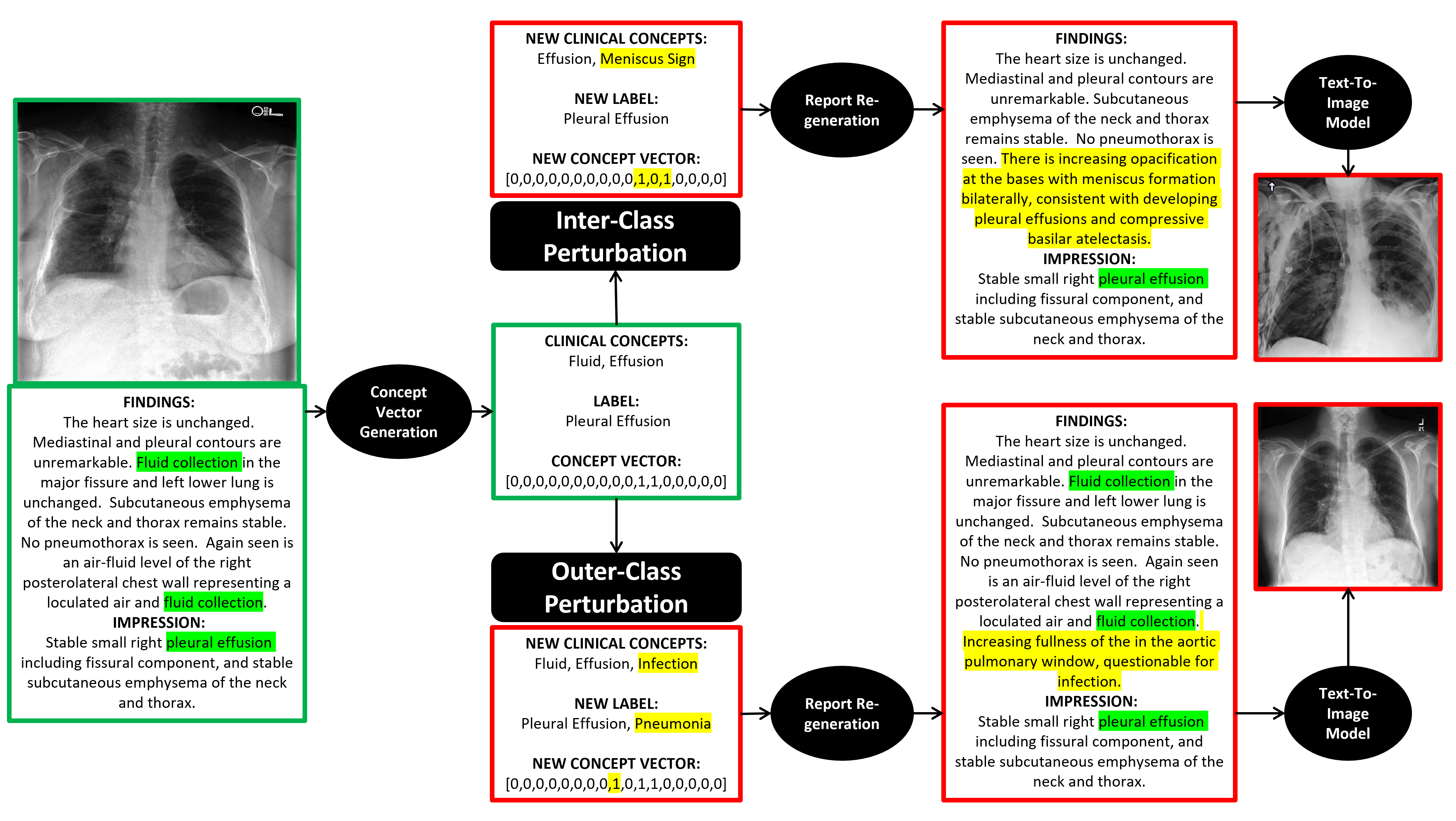}}
\caption{Visualisation of the CoRPA pipeline. A chest X-ray and corresponding report are used to generate a concept vector. Green-highlighted phrases are used by the algorithm (see Table~\ref{tab:concepts}). The image-report pair is labelled as Pleural Effusion as the \texttt{Effusion} and \texttt{Fluid} concepts are present. The concept vector is perturbed four times - two inter-class and two outer-class perturbations. We show one perturbation of each type for visibility. Adversarial reports are generated through sentence manipulation based on these perturbed vectors. Sentences related to removed concepts (\texttt{Fluid} in the inter-class example) are removed from the report. A new sentence for the added concepts (\texttt{Meniscus Sign} in the inter-class example, and \texttt{Infection} in the outer-class example) is inserted (yellow). Reports are then input into a text-to-image Stable Diffusion model to produce an adversarial image.}
\label{fig:pipeline}
\end{figure*}

This section presents CoRPA (Concept-based Report Perturbation Attack), a clinically-focused, untargeted black-box adversarial methodology that uses a text-to-image Stable Diffusion model to generate adversarial medical images from adversarial radiological reports, generated through perturbations to clinical concept vectors. The CoRPA pipeline for our dataset of chest X-rays and associated reports (Figure~\ref{fig:pipeline}) is as follows:

\begin{enumerate}
    \item For each chest X-ray - report pair, a clinical concept vector is generated. This vector captures the presence of pre-defined clinical concepts within the report text, accounting for contextual factors such as negations. See Section~\ref{labelling}.
    \item Four random perturbations of the concept vector are created: two inter-class perturbations and two outer-class perturbations, or four outer-class perturbations in the case of single-concept classes (Healthy, Cardiomegaly).
    \item Adversarial reports are reconstructed using the perturbed concept vectors.
    \item Reconstructed reports are input into a text-to-image Stable Diffusion model, pretrained on the MIMIC-CXR-JPG dataset, to generate an adversarial image for each report.
\end{enumerate}

This approach yields four adversarial images per test set image, resulting in a total of 14,360 adversarial images. The choice of four perturbations is customizable and was made to balance dataset size and diversity while maintaining computational feasibility. 

\subsection{Concept Vector Perturbations}

We propose two types of concept vector perturbations, each designed to simulate realistic medical scenarios involving adversarial inputs, primarily arising from interpretative challenges in radiological reports for chest X-rays. Chest radiography, despite being the most commonly performed imaging examination globally, remains susceptible to frequent interpretation errors \cite{b66}. By implementing these perturbations, we aim to capture such behaviours systematically.

Inter-class perturbations modify only the concepts associated with the pathology label of the chest X-ray. For instance, as illustrated in Figure~\ref{fig:pipeline}, a chest X-ray labelled with Pleural Effusion originally characterized by the concepts \texttt{Fluid} and \texttt{Effusion} could be perturbed to instead include \texttt{Effusion} and \texttt{Meniscus Sign}. This approach reflects scenarios where radiologists identify alternative but valid indicators of the same pathology, and accommodates for variations in the descriptive language used by different radiologists for similar findings.

Outer-class perturbations introduce concepts from a second, randomly selected pathology into the chest X-ray. For example, as shown in Figure~\ref{fig:pipeline}, a chest X-ray labelled with Pleural Effusion might be augmented with the concept \texttt{Infection}, which belongs to the Pneumonia pathology class. This perturbation type emulates situations where radiologists misdiagnose a condition, a phenomenon well-documented in the literature due to the notable rates of false positives and negatives in chest X-ray interpretations \cite{b67} \cite{b68}.

We perturb concept vectors using the following algorithm. All random generations use Python’s random package with seed 2. Concept vectors are binary, where 1 denotes the presence of a concept and 0 indicates its absence. A visual example of the CoRPA pipeline, including concept vector perturbations, is shown in Figure~\ref{fig:pipeline}.
\newline

\begin{figure}[tb]
\centerline{\includegraphics[width=
0.9\linewidth]{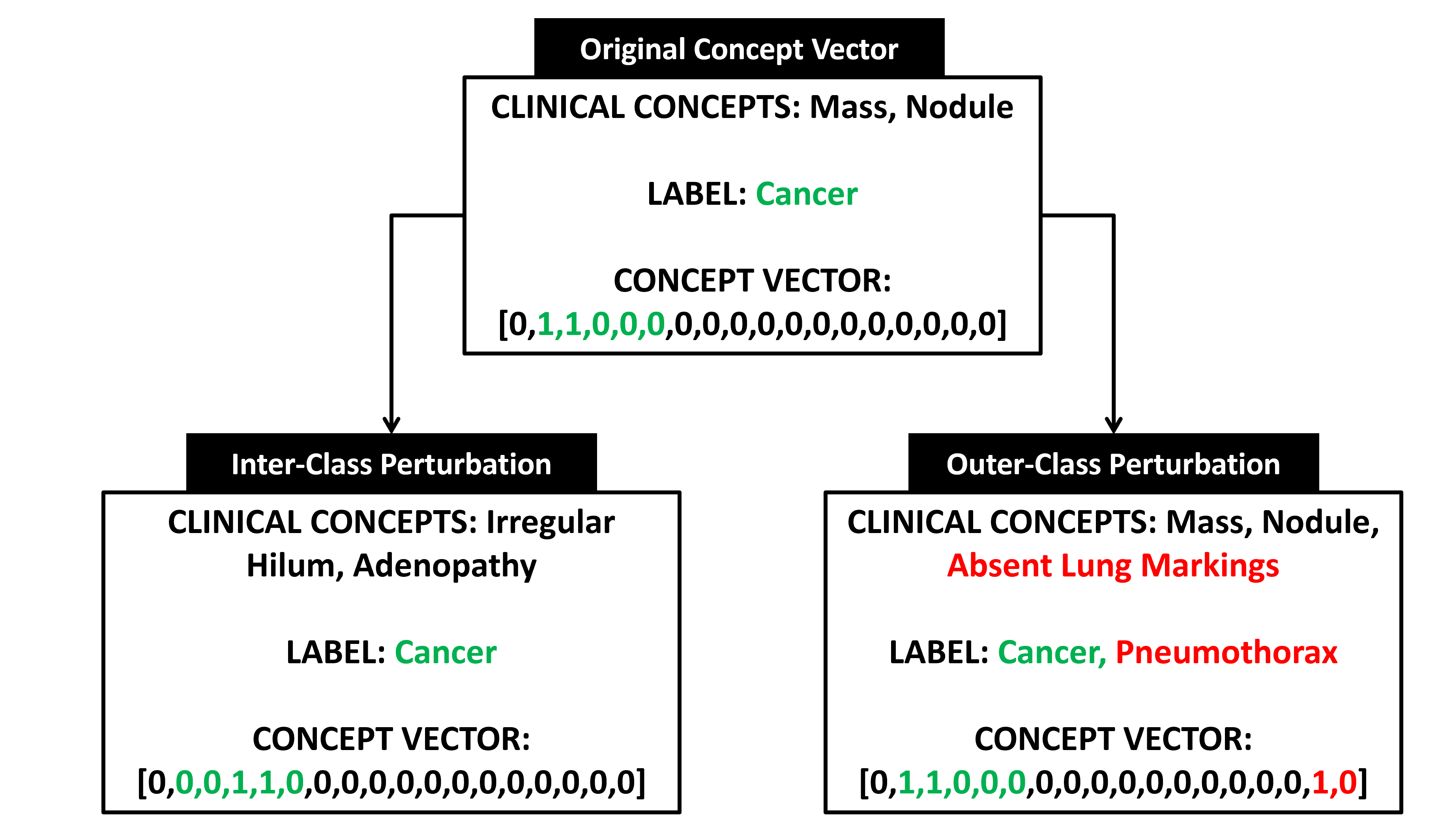}}
\caption{Visualization of an inter- and outer-class perturbation for a cancerous concept vector. For inter-class perturbations, only the concepts relating to the original class (green) are perturbed. For outer-class perturbations, concepts relating to the original class (green) remain the same, and a random perturbation of the concepts relating to the randomly selected new class (red) is generated.}
\label{fig:perturb}
\end{figure}

\noindent Input:
\begin{itemize}
    \item $V$: Binary concept vector corresponding to a chest X-ray labelled with a given class $L$.
    \item $K\textsubscript{inter}$ = 2: Number of inter-class perturbations.
    \item $K\textsubscript{outer}$ = 2: Number of outer-class perturbations.
\end{itemize}
\noindent Output:
\begin{itemize}
    \item ${V^\prime}\textsubscript{inter}$: Set of $K\textsubscript{inter}$ inter-class perturbed vectors.
    \item ${V^\prime}\textsubscript{outer}$: Set of $K\textsubscript{outer}$ outer-class perturbed vectors.\newline
\end{itemize}
Note that classes Healthy and Cardiomegaly have only one clinical concept, and therefore inter-class perturbations are impossible. In these cases, we set $K\textsubscript{inter}$ = 0 and $K\textsubscript{outer}$ = 4. Examples of both inter-class and outer-class concept vector perturbations are shown in Figure~\ref{fig:perturb}.

To generate inter-class perturbations, we first identify the concepts within vector $V$ which correspond to class $L$. We generate perturbed vectors by randomly modifying only these elements of $V$, leaving the rest of the vector unchanged. The perturbed vector must satisfy two conditions: it must differ from the original vector, and at least one concept corresponding to class $L$ must remain active. Each perturbed vector is validated to ensure that it belongs to the valid concept vector list for class $L$ based on the dataset, before it is added to ${V^\prime}\textsubscript{inter}$. This process is repeated until $K\textsubscript{inter}$ valid perturbations are obtained.

For outer-class perturbations, we leave the elements of $V$ corresponding to class $L$ unchanged, and instead randomly select a new target class $L^\prime$. Perturbed vectors are generated by randomly modifying the elements of $V$ associated with class $L^\prime$, ensuring that at least one concept corresponding to class $L^\prime$ is active. Each perturbed vector is again validated to ensure that it belongs to the valid concept vector list for class $L^\prime$ before it is added to ${V^\prime}\textsubscript{outer}$, and the process is repeated until $K\textsubscript{outer}$ valid perturbations are created.

\begin{figure*}[tb]
\centerline{\includegraphics[width=
\linewidth]{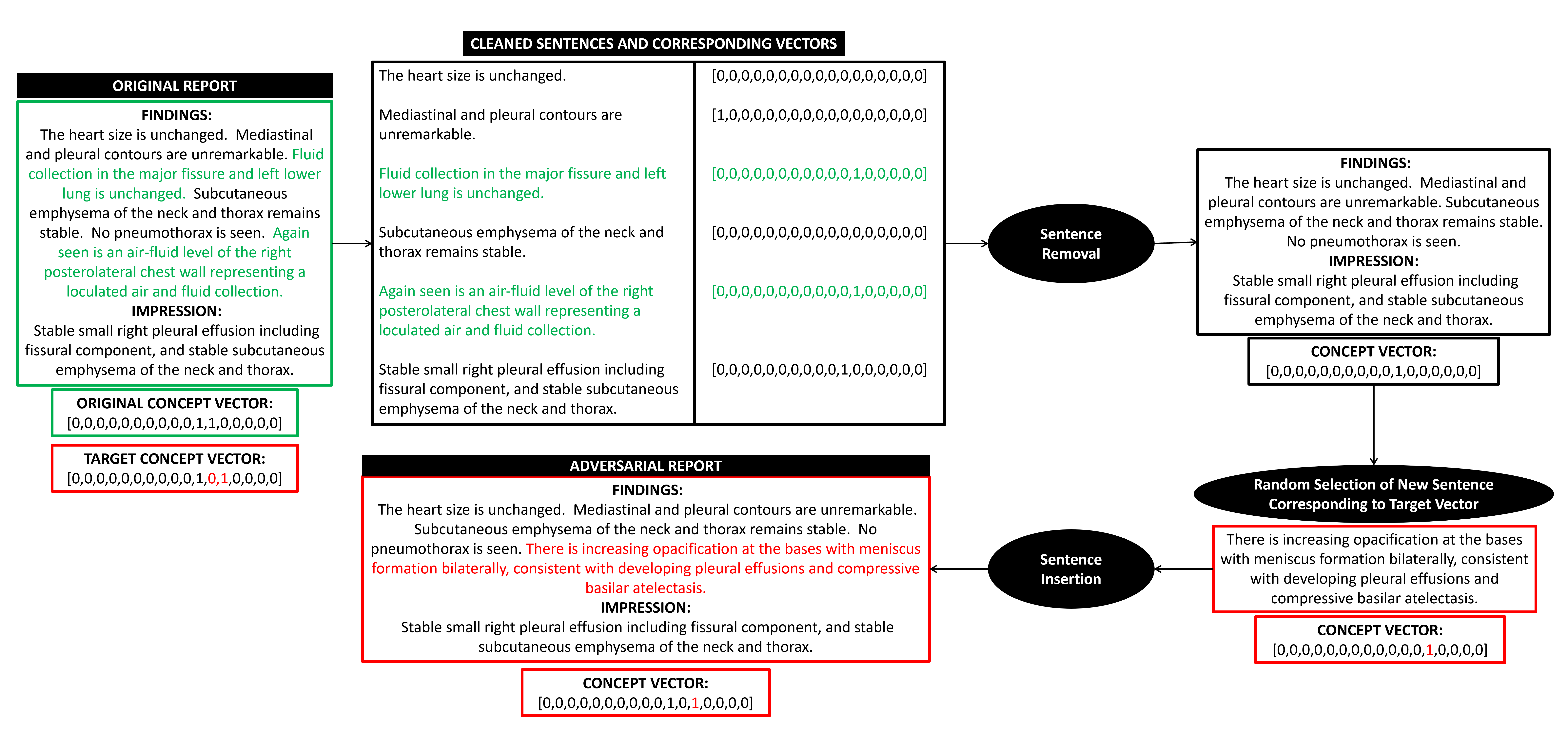}}
\caption{Visualisation of the re-generation of adversarial radiological reports from perturbed concept vectors through sentence manipulation, for an inter-class perturbation example. The sentences contributing to the original concept \texttt{Fluid} which is not present in the perturbed target vector (green) are removed from the text. Note this only applies to sentences within the 'cleaned' list, as defined in Section~\ref{labelling}. A random sentence corresponding to the new concept \texttt{Meniscus Sign} in the target concept vector is then chosen from the concept-to-sentence mapping (red), and inserted into the report. The sentence corresponding to the Healthy class is ignored due to the presence of pathology-indicative concepts.}
\label{fig:sents}
\end{figure*}

\subsection{Adversarial Report Generation}

The perturbed concept vectors are then reconstructed into radiological reports. We first generate a sentence-to-concept mapping. For each concept $c$ in our predefined set of 17 clinical concepts, we identify five unique sentences, randomly chosen from the original test set reports, where only $c$, or $c$ along with one additional concept, are set to 1 in the corresponding concept vector. 

This mapping facilitates the insertion and removal of sentences within a report, enabling the generation of adversarial reports that align with the target perturbed concept vectors. The following process is executed for each perturbed concept vector produced in the previous CoRPA step. We retain a mapping file linking perturbed vectors to their respective original reports to allow for correct data linking. An illustrative example of adversarial report generation, with the inserted and removed sentences highlighted, is presented in Figure~\ref{fig:sents}.
\newline

\noindent Input:
\begin{itemize}
    \item $R$: Original radiological report
    \item $V$: Original concept vector for $R$
    \item $P$: Perturbed concept vector generated from $V$
\end{itemize}
\noindent Output:
\begin{itemize}
    \item $R^\prime$: Adversarial report generated from sentence manipulation. \newline
\end{itemize}

Initially, the original report $R$ is converted into a list of cleaned sentences, and the corresponding concept vectors for each sentence are computed as detailed in Section~\ref{labelling}. This list, comprising the report sentences and their associated concept vectors, enables us to identify which sentences in the original report contribute to the concepts in vector $V$ that are set to 1. In the case of inter-class perturbations, the sentences contributing to concepts which are set to 1 in the original vector $V$ but not the target perturbed vector $P$, are removed from the report. In both perturbation cases, a sentence which corresponds to the target perturbed concept, which is set to 1 in vector $P$, is then randomly selected from the previously defined sentence-to-concept mapping. This selected sentence is inserted into the report, specifically in the second-to-last position, as the final sentence in the report is typically reserved for a summary statement.

%\noindent Report Generation Method:

%\begin{enumerate}
%    \item Report Conversion: Convert $R$ into a list of cleaned sentences $S$, and calculate the concept vectors for each of these sentences (see Section~\ref{labelling}). 
%    \item Sentence Removal: For each concept in $V$ set to 1, determine which sentences in $S$ need to be removed in order to set those concepts to 0. Remove the identified sentences from $S$.
%    \item Sentence Insertion: Randomly select sentences from the sentence-to-concept mapping $M$ that correspond to the concepts required to generate the goal vector $P$. Insert the selected sentences into $S$ in the second-to-last position (the last sentence is typically a summary statement).
%    \item Report Formatting: Combine the modified sentence list $S$ back into a string to form the adversarial report $R^\prime$.
%    \item Validation: Verify that $R^\prime$ corresponds to the goal perturbed concept vector $P$.
%\end{enumerate}

\subsection{Adversarial Image Generation}

For each chest X-ray-report pair in our test set, four adversarial reports are generated with inter or outer class perturbations. These reports are processed through a text-to-image generative model to produce corresponding adversarial images, which will be used to assess model robustness. 

Adversarial reports are cleaned using the approach outlined in Section~\ref{labelling} to ensure only the relevant clinical information is being passed to the generative model. These cleaned adversarial reports are directly used as model prompts.

We used a generative Stable Diffusion model, refined using the LoRA (Low-Rank Adaptation) \cite{b62} fine-tuning approach on the MIMIC-CXR-JPG dataset. Stable Diffusion models have recently demonstrated an ability to generate high-quality images from diverse text inputs, while being more computationally efficient than other generative models \cite{b44}. We fine-tuned this model using preexisting weights derived from the MIMIC-CXR-JPG dataset, publicly available on Huggingface \cite{b63}. These weights were originally created for PIE, a high-performance generative model focused on disease progression in chest X-rays \cite{b64}. We fine-tune our Stable Diffusion model using the LoRA technique based on its ability to sustain high performance with minimal resource consumption. LoRA works by introducing trainable low-rank matrices into model weights, instead of altering entire network architectures.

\begin{table*}[t]
\caption{Performance of Model Architectures by Class on the \textbf{Original Test Set}. We present Average Precision (AP) and AUROC (AUC) scores for each class, as well as the mean AP (mAP) and mean AUROC score for each model. Best performances are shown in \textbf{bold}.}
\begin{center}
\begin{tabular}{|c|cc|cc|cc|cc|cc|cc|cc|}
\hline
 & \multicolumn{2}{c|}{\textbf{ResNet50}} &  \multicolumn{2}{c|}{\textbf{ResNet101}} & \multicolumn{2}{c|}{\textbf{ResNeXt101}} &  \multicolumn{2}{c|}{\textbf{DenseNet161}} & \multicolumn{2}{c|}{\textbf{ConvNeXt-S}} &  \multicolumn{2}{c|}{\textbf{ConvNeXt-B}} &  \multicolumn{2}{c|}{\textbf{EfficientNetV2-S}} \\
 & \textbf{AP} & \textbf{AUC$^{\mathrm{*}}$}  & \textbf{AP} & \textbf{AUC$^{\mathrm{*}}$} & \textbf{AP} & \textbf{AUC$^{\mathrm{*}}$} & \textbf{AP} & \textbf{AUC$^{\mathrm{*}}$} & \textbf{AP} & \textbf{AUC$^{\mathrm{*}}$} & \textbf{AP} & \textbf{AUC$^{\mathrm{*}}$} & \textbf{AP} & \textbf{AUC$^{\mathrm{*}}$} \\
\hline
%['Cancer', 'Cardiomegaly', 'Effusion', 'Healthy', 'Pneumonia', 'Pneumothorax']
Healthy & 0.891 & 0.922 & 0.892 & 0.926 & 0.894 & 0.925 & \textbf{0.907} & \textbf{0.936} & 0.799 & 0.882 & 0.739 & 0.883 & \textbf{0.907} & 0.935 \\
Cancer & 0.155 & 0.645 & 0.148 & 0.660 & 0.161 & 0.708 & 0.163 & 0.692 & 0.145 & 0.626 & 0.142 & 0.618 & \textbf{0.177} & \textbf{0.736} \\
Cardiomegaly & 0.444 & 0.858 & 0.435 & 0.839 & 0.451 & 0.865 & \textbf{0.478} & \textbf{0.882} & 0.333 & 0.673 & 0.283 & 0.627 & 0.468 & 0.870 \\
Pleural Effusion & 0.521 & 0.882 & 0.524 & 0.882 & 0.525 & 0.876 & \textbf{0.548} & 0.895 & 0.378 & 0.687 & 0.292 & 0.558 & 0.531 & \textbf{0.899} \\
Pneumonia & 0.229 & 0.710 & 0.248 & 0.742 & 0.254 & 0.772 & \textbf{0.289} & \textbf{0.781} & 0.192 & 0.607 & 0.183 & 0.565 & 0.264 & 0.768 \\
Pneumothorax & 0.259 & 0.860 & 0.224 & 0.852 & 0.203 & 0.841 & \textbf{0.329} & 0.874 & 0.158 & 0.681 & 0.116 & 0.668 & 0.301 & \textbf{0.879} \\
\hline
Mean & 0.417 & 0.813 & 0.412 & 0.817 & 0.415 & 0.831 & \textbf{0.452} & 0.843 & 0.334 & 0.693 & 0.293 & 0.653 & 0.441 & \textbf{0.848} \\
\hline
\multicolumn{8}{l}{$^{\mathrm{*}}$We denote AUROC as "AUC" to conserve space.}
\end{tabular}
\label{tab:orig_results}
\end{center}
\end{table*}

\begin{table*}[t]
\caption{Performance of Model Architectures by Class on the \textbf{CoRPA Adversarial Test Set}. We present Average Precision (AP) and AUROC (AUC) scores for each class, as well as the mean AP (mAP) and mean AUROC score for each model. We also show the decrease in mAP and mAUROC scores compared to the original test set (Mean Decrease). Best performances are shown in \textbf{bold}.}
\begin{center}
\begin{tabular}{|c|cc|cc|cc|cc|cc|cc|cc|}
\hline
 & \multicolumn{2}{c|}{\textbf{ResNet50}} &  \multicolumn{2}{c|}{\textbf{ResNet101}} & \multicolumn{2}{c|}{\textbf{ResNeXt101}} &  \multicolumn{2}{c|}{\textbf{DenseNet161}} & \multicolumn{2}{c|}{\textbf{ConvNeXt-S}} &  \multicolumn{2}{c|}{\textbf{ConvNeXt-B}} &  \multicolumn{2}{c|}{\textbf{EfficientNetV2-S}} \\
 & \textbf{AP} & \textbf{AUC$^{\mathrm{*}}$}  & \textbf{AP} & \textbf{AUC$^{\mathrm{*}}$} & \textbf{AP} & \textbf{AUC$^{\mathrm{*}}$} & \textbf{AP} & \textbf{AUC$^{\mathrm{*}}$} & \textbf{AP} & \textbf{AUC$^{\mathrm{*}}$} & \textbf{AP} & \textbf{AUC$^{\mathrm{*}}$} & \textbf{AP} & \textbf{AUC$^{\mathrm{*}}$} \\
\hline
%['Cancer', 'Cardiomegaly', 'Effusion', 'Healthy', 'Pneumonia', 'Pneumothorax']
Healthy & 0.431 & 0.629 & 0.428 & \textbf{0.679} & 0.427 & 0.670 & 0.431 & 0.611 & 0.426 & 0.651 & \textbf{0.436} & 0.640 & \textbf{0.436} & 0.655\\
Cancer & 0.189 & 0.493 & \textbf{0.196} & 0.513 & 0.185 & 0.482 & 0.192 & 0.498 & 0.187 & 0.493 & 0.182 & 0.478 & 0.194 & \textbf{0.514} \\
Cardiomegaly & 0.273 & 0.532 & 0.275 & 0.516 & 0.266 & 0.521 & 0.261 & 0.511 & 0.251 & 0.485 & 0.252 & 0.484 & \textbf{0.283} & \textbf{0.542} \\
Pleural Effusion & 0.266 & 0.532 & 0.278 & 0.535 & 0.272 & 0.536 & 0.273 & 0.549 & 0.267 & 0.530 & 0.285 & 0.544 & \textbf{0.304} & \textbf{0.588} \\
Pneumonia & 0.184 & 0.531 & 0.159 & 0.470 & 0.179 & 0.491 & 0.148 & 0.446 & 0.177 & 0.507 & \textbf{0.204} & \textbf{0.557} & 0.165 & 0.474 \\
Pneumothorax & 0.105 & 0.517 & 0.101 & 0.510 & 0.115 & \textbf{0.536} & \textbf{0.120} & 0.531 & 0.097 & 0.491 & 0.094 & 0.490 & 0.101 & 0.511 \\
\hline
Mean & 0.241 & 0.539 & 0.240 & 0.537 & 0.241 & 0.539 & 0.238 & 0.524 & 0.234 & 0.526 & 0.242 & 0.532 & \textbf{0.247} & \textbf{0.547} \\
\hline
Mean Decrease & 0.176 & 0.274 & 0.172 & 0.280 & 0.174 & 0.292 & \textbf{0.214} & \textbf{0.319} & 0.100 & 0.167 & 0.051 & 0.121 & 0.194 & 0.301 \\
\hline
\multicolumn{8}{l}{$^{\mathrm{*}}$We denote AUROC as "AUC" to conserve space.}
\end{tabular}
\label{tab:corpa_results}
\end{center}
\end{table*}

\section{Results}

In this section, we evaluate CoRPA through two distinct approaches. First, we assess the classification performances of seven model architectures on both the original test set of 3590 chest X-rays, and the adversarial test set of 14,360 images generated by CoRPA. We evaluate this through Mean Average Precision (mAP) and Mean Area Under the Receiver Operating Characteristic Curve (mAUROC). Second, we compare CoRPA's performance against three other widely-used adversarial attacks in medical image classification tasks, focusing on the Attack Success Rate (ASR). We aim to provide a comprehensive understanding of CoRPA's effectiveness as a clinically-focused adversarial attack.

In the case of CoRPA’s inter-class perturbations, the adversarial images are generated by modifying clinical concepts within the same pathology class as the original image. Since these perturbations do not introduce features indicative of a different pathology, the model is considered robust and correct if the predicted label remains unchanged. In contrast, outer-class perturbations introduce clinical concepts associated with a different pathology class, thereby creating adversarial images with features of both the original and the newly introduced pathology. For such cases, the model is deemed robust and correct if both the original and newly introduced labels are the top two highest-scoring predicted classes. This is because the adversarial image retains the defining characteristics of the original class, and should therefore be recognised as such, even as it incorporates features of the second class, which may slightly alter the model's predictions.

\begin{figure}[tb]
\centering
\begin{subfigure}{0.48\textwidth}
    \includegraphics[width=\textwidth]{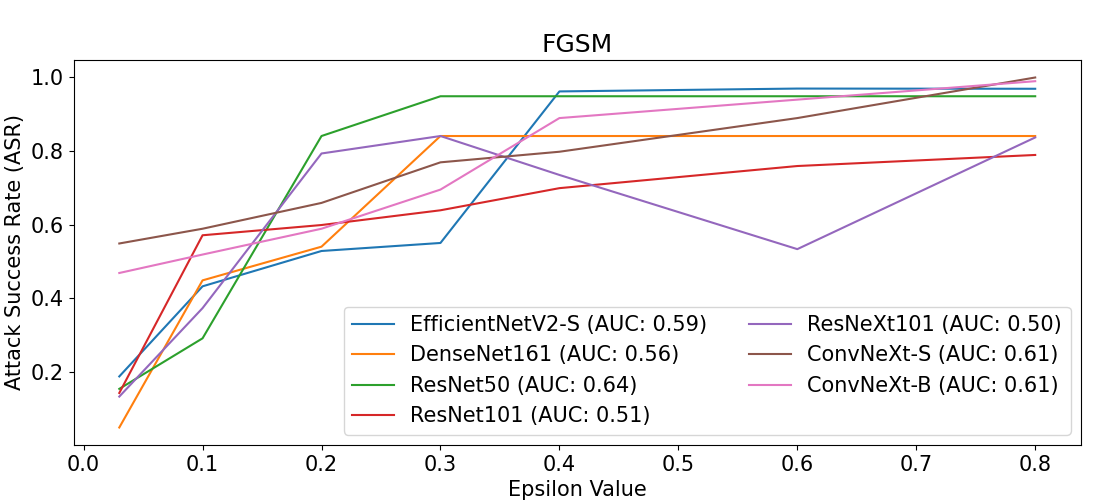}
    \subcaption{FGSM}
\end{subfigure}
\begin{subfigure}{0.48\textwidth}
    \includegraphics[width=\textwidth]{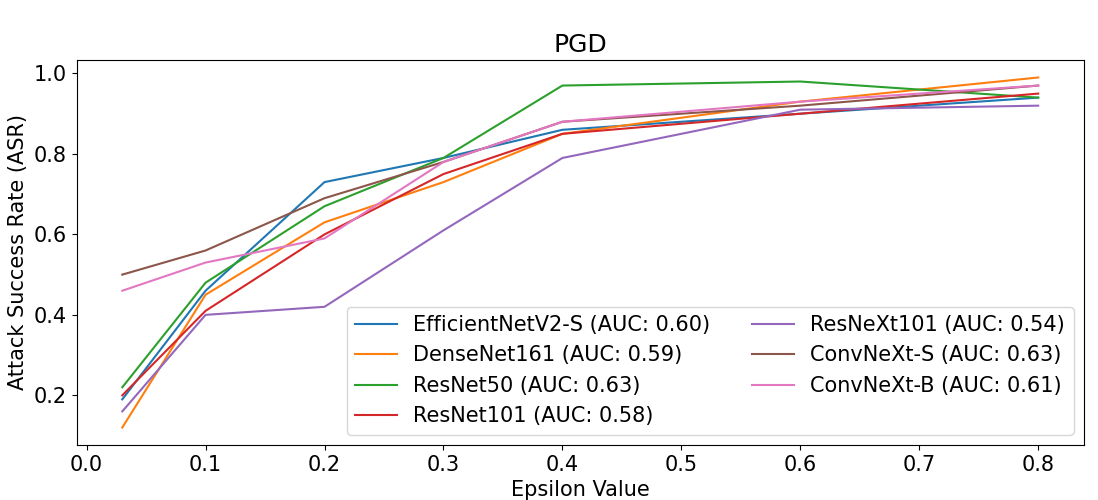}
    \subcaption{PGD}
\end{subfigure}
\begin{subfigure}{0.48\textwidth}
    \includegraphics[width=\textwidth]{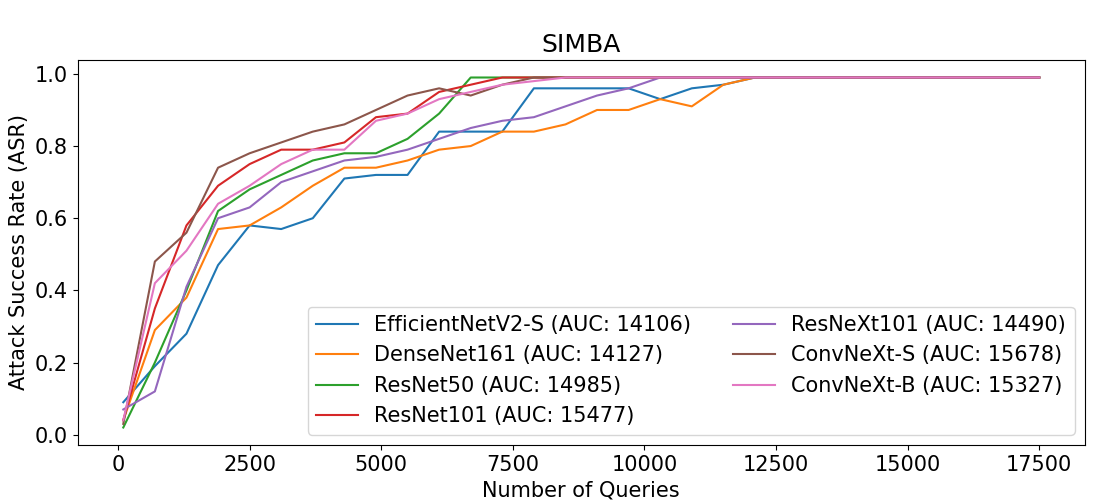}
    \subcaption{SimBA}
\end{subfigure}
\caption{Attack Success Rate (ASR) curves for \textbf{FGSM}, \textbf{PGD}, and \textbf{SimBA} across all tested models. The legends display the Area Under the Curve (AUC) values, where higher AUCs correspond to steeper curves, indicating lower robustness of the models to the attack.}
%\vspace{-10pt}
\label{fig:asrs}
\end{figure}

\subsection{Model Performance}

Following the MICCAI CXR-LT challenge \cite{b54}, whose model backbones are evaluated in this study, we assess the classification performance of these seven models using Mean Average Precision (mAP) and Mean AUROC (mAUROC). Given the imbalanced nature of our dataset, we anticipate high mAUROC values but relatively low mAP values. For reference, in the significantly more imbalanced dataset used in the CXR-LT challenge, the top-performing model achieved a mAP score of 0.372 \cite{b60}.

The classification performance of each model on the original test set of 3,590 chest X-rays is summarized in Table~\ref{tab:orig_results}. We report the Average Precision (AP) and AUROC for each class, along with the overall mAP and mAUROC for each model. DenseNet161 achieved the highest AP, while the best AUROC was achieved by EfficientNetV2-S. Both ConvNeXt models exhibited poor performance across both metrics. 

The performance of each model on the adversarial test set generated by CoRPA is presented in Table~\ref{tab:corpa_results}. This table also presents the difference in mAP and mAUROC scores between the original and adversarial test sets for each model. EfficientNetV2-S outperformed other models on this test set in terms of both AP and AUROC, despite a relatively high decrease in performance regarding AUC between datasets. Both ConvNeXt models demonstrated the poorest performance, consistent with the original test set, as well as the smallest mean decrease in performance. Notably, DenseNet161 displayed the most dramatic performance decrease between the two test sets. Generally, CoRPA has the biggest effect on model performance for the Healthy class.

\begin{table}[tb]
\caption{Attack Success Rate (ASR) of \textbf{CoRPA} for all models. Results are presented for all adversarial examples collectively, as well as separately for those generated through inter-class and outer-class concept vector perturbations. Higher ASR indicates lower robustness. }
\begin{center}
\begin{tabular}{|c|c|c|c|}
\hline
\textbf{Model} & \textbf{Inter-Class} & \textbf{Outer-Class} & \textbf{All Perturbations}\\
\hline
%['Cancer', 'Cardiomegaly', 'Effusion', 'Healthy', 'Pneumonia', 'Pneumothorax']
ResNet50 & 0.385 & 0.855 & 0.761 \\
ResNet101 & 0.433 & 0.860 & 0.775 \\
ResNeXt101 & 0.554 & 0.885 & 0.819 \\
DenseNet161 & 0.536 & 0.886 & 0.816 \\
ConvNeXt-S & 0.430 & 0.902 & 0.808 \\
ConvNeXt-B & 0.419 & 0.900 & 0.804 \\
EfficientNetV2-S & 0.553 & 0.889 & 0.822 \\
\hline
Mean & 0.473 & 0.882 & 0.801 \\
\hline
\end{tabular}
\label{tab:corpa_asr_results}
\end{center}
\end{table}

\subsection{Comparison with Other Attacks}

To further investigate the robustness of each model and the performance and utility of CoRPA as a clinically focused black-box attack, we calculate the Attack Success Rate (ASR) of CoRPA and compare it to other widely-used adversarial attacks. ASR is defined as the proportion of adversarial examples that successfully cause the model to misclassify the input. A higher ASR indicates lower robustness of the model against the attack.

We compare CoRPA’s performance to three other attacks: SimBA (black-box), FGSM, and PGD (white-box). These attacks are typically evaluated in the literature through ASR curves, where ASR is calculated over a range of values for their respective variable parameters. For FGSM and PGD, the variable is the epsilon value, which controls the magnitude of perturbations \cite{b65}. SimBA, a score-based black-box attack that iteratively queries the model to modify adversarial perturbations, uses the number of queries as its variable. The ASR curves for these three attacks across each of our models are shown in Figure~\ref{fig:asrs}. We present the Area Under the Curve (AUC) within the legends of each sub-figure. Higher AUC values correspond to steeper curves, indicating lower robustness of the models to the attack.

Since CoRPA does not involve a variable parameter, it does not generate an ASR curve. Instead, we present its ASR values in Table~\ref{tab:corpa_asr_results}. These values are shown for the entire adversarial test set, as well as for the subsets of adversarial examples generated by inter-class or outer-class concept vector perturbations.

As expected, the ASR of outer-class adversarial examples in Table~\ref{tab:corpa_asr_results} is higher than that of inter-class examples. Outer-class examples introduce clinical features of a second pathology class, effectively combining characteristics of both the original and the newly introduced pathology. This results in more significant differences between the adversarial images and the original images compared to inter-class perturbations, that only alters features within the same pathology class. These larger differences and the need to detect both the original and second newly added pathology make the model more likely to misclassify, as it encounters features indicative of multiple pathologies that may confuse its predictions.

Across the seven model architectures evaluated, CoRPA achieves an ASR of 88.2\% for outer-class adversarial chest X-rays, and 47.3\% for inter-class. The adversarial dataset generated by CoRPA contains only 20\% inter-class perturbations, due to the presence of classes with a single clinical concept (Healthy, Cardiomegaly). As a result, the ASR on the full adversarial test set remains high - 80.1\% across all models.

The models demonstrating the lowest robustness to CoRPA, based on their high ASR scores, are EfficientNetV2-S, ResNeXt101, and DenseNet161. Notably, these same models exhibit the highest robustness to SimBA, as indicated by the low AUC scores of its ASR curve (Figure~\ref{fig:asrs}). Additionally, ResNeXt101 and DenseNet161 rank among the three most robust models against FGSM and PGD, alongside ResNet101.

This observation suggests that while models may exhibit resilience to adversarial examples generated by general-purpose attacks, they can remain vulnerable to adversarial examples specifically crafted for clinical contexts. This highlights the critical need for CoRPA as a tool to evaluate model robustness in scenarios that are more reflective of real-world clinical challenges.

\section{Conclusion}

In this study, we introduced the Concept-based Report Perturbation Attack (CoRPA), a clinically focused black-box adversarial attack framework designed to mimic real-world clinical scenarios that could lead to diagnostic errors, such as missed diagnoses, clinical variability, and misinterpretation. By leveraging clinical concepts in radiology reports and generative models, CoRPA generates realistic adversarial chest X-ray images and radiological reports that exploit domain-specific vulnerabilities in deep learning models.

Our evaluation on the MIMIC-CXR-JPG dataset revealed that CoRPA exposes significant gaps in the robustness of several state-of-the-art deep learning architectures, which otherwise exhibit strong resilience to conventional adversarial attacks. This demonstrates the necessity of domain-specific adversarial testing for medical AI systems, as traditional approaches may fail to reflect real-world clinical mistakes. This work lays a foundation for future research in clinically focused robustness testing, contributing to safer and more reliable AI deployments in medical diagnostics.

\subsection{Limitations and Future Work}

The primary limitations of CoRPA include its dependence on medical datasets with both images and corresponding radiological reports, as well as the need for clinician input when defining the clinical concept space for a specific dataset. However, our consultant radiologist was able to identify the clinical concepts listed in Table~\ref{tab:concepts} in under an hour, and once these concepts are established, no further clinician involvement is necessary for the CoRPA pipeline. %This study focuses solely on adversarial attack methods. 
Future research will focus on developing robust defense mechanisms that integrate clinical semantics and domain-specific features to enhance model resilience against clinically relevant attacks, such as adversarial training with CoRPA-generated images. We also intend to support other medical imaging modalities, such as MRIs and CT scans.

%%%%%%%%%%%%%%%%%%%%%%%%%%%%%%%%%%%%%%%%%%%
%%%%%%%%%%%%%%%%%%%%%%%%%%%%%%%%%%%%%%%%%%%
%%%%%%%%%%%%%%%%%%%%%%%%%%%%%%%%%%%%%%%%%%%
%%%%%%%%%%%%%%%%%%%%%%%%%%%%%%%%%%%%%%%%%%%

\end{document}